\documentclass[%
 aip,
 amsmath,amssymb,
 reprint,%
]{revtex4-1}

\usepackage{graphicx}% Include figure files
\usepackage{dcolumn}% Align table columns on decimal point
\usepackage{bm}% bold math
\usepackage[utf8]{inputenc}
\usepackage[T1]{fontenc}
\usepackage{mathptmx}
\usepackage{etoolbox}
\usepackage{tikz}
\makeatletter
\def\@email#1#2{%
 \endgroup
 \patchcmd{\titleblock@produce}
  {\frontmatter@RRAPformat}
  {\frontmatter@RRAPformat{\produce@RRAP{*#1\href{mailto:#2}{#2}}}\frontmatter@RRAPformat}
  {}{}
}%
\makeatother
\begin{document}

\preprint{AIP/123-QED}

\title{Exploring the Nexus between Thermodynamic Phase Transitions and Geometric Fractals through Systematic Lattice Point Classification}
% Force line breaks with \\
\author{Yonglong Ding}
\affiliation{School of Physics, Beijing Institute of Technology, Beijing 100081, China}
\affiliation{Beijing Computational Science Research Center, Beijing 100193, China}
%\author{B. Author}%
 \email{ylding@csrc.ac.cn.}
%\affiliation{ 
%Authors' institution and/or address%\\This line break forced with \textbackslash\textbackslash
%}%

% \homepage{http://www.Second.institution.edu/~Charlie.Author.}
%\affiliation{%
%Second institution and/or address%\\This line break forced% with \\
%}%

\date{\today}% It is always \today, today,

\begin{abstract}
Fractals are ubiquitous in the natural world, and their connection with phase transitions has been widely observed. This study investigates mechanisms of fractal formation from the perspective of phase transitions. A novel set of probability calculation methods is introduced to establish a direct link between fractals and phase transitions. Notably, in the Ising model, a specific category of boundary lattice points undergoes a phase transition when the associated weight reaches approximately 0.4. The identified correlation between phase transitions and fractals suggests the emergence of fractal structures at this critical weight. The paper offers supporting evidence for this conclusion through the deliberate manipulation of the proposed probability-based method. This research contributes to a deeper understanding of the interplay between fractals and phase transitions, providing valuable insights for further exploration in diverse scientific domains.
\end{abstract}

\maketitle

\begin{quotation}
Fractals represent visually stunning phenomena in nature, with potential existence at phase transition points where the association between phase transitions and fractal structures remains inadequately comprehended. Establishing a direct relationship between fractals and phase transitions could serve as a pivotal catalyst for a deeper comprehension of both phenomena.
\end{quotation}

\section{\label{sec:level1}Introduction}

To comprehend fractal structures from the standpoint of phase transitions, it is imperative to construct a well-suited framework of computational methodologies designed for the analysis of phase transitions. The process of phase transition is remarkably intricate from a computational perspective, with direct calculations demanding staggering computational resources. Research on phase transitions commonly employs algorithms such as Monte Carlo simulations\cite{RevModPhys.73.33,sarrut2021advanced,luo2022hybrid}, tensor networks\cite{orus2019tensor,pan2022simulation,gray2021hyper,yuan2021quantum}, and scaling laws\cite{shao2016quantum,halperin2019theory,manipatruni2019scalable}. Landau's mean-field theory offers elegant conclusions, particularly for higher-dimensional Ising models. Monte Carlo algorithms exhibit high computational efficiency\cite{tang2018role,carlson2015quantum,lynn2019quantum,nemeth2021stochastic,rioux2022monte,andersen2019practical}. There exists a certain mapping relationship between the Ising model and computational methods like the XY model\cite{barouch1970statistical,kosterlitz1974critical}, JQ model\cite{desai2020first}, and Potts model\cite{wu1982potts}. Through the computations of these models, a wide range of physical properties can be revealed.

Nevertheless, these approaches have required extensive computational resources. With the emergence of more intricate models, utilizing simulations to determine critical points has become increasingly challenging. Monte Carlo algorithms encounter sign problems\cite{henelius2000sign,alexandru2022complex}, and efforts have been made to address this issue\cite{mondaini2022quantum,zhang2022fermion,berger2021complex}.Even in the case of simple interaction models, there exist rich physical properties, with symmetry playing a crucial role\cite{kong2020algebraic,hu2020symmetry}. Presently, the acceleration of this process is achieved through multiple approaches, finding extensive applications. The solutions to these optimization problems can be mainly categorized into three classes: first, through algorithmic simulations; second, by combining algorithmic simulations with hardware acceleration\cite{ortega2015fpga,yang2019high,preis2009gpu}; third, utilizing physical annealing\cite{suzuki2007quantum}. The study of the Ising model has also given rise to various algorithms and theories\cite{dziarmaga2005dynamics,baxter1978399th}. The exploration of the Ising model continues to yield diverse physical phenomena\cite{zhang2020loop}. High-dimensional Ising models have consistently captivated researchers' exploration efforts\cite{fang2023geometric,park2022bayesian,kryzhanovsky2021analytical,dagan2021learning}.

Building upon the groundwork laid by previous studies on phase transitions, this paper introduces a novel methodology relying on probabilistic computations. This approach facilitates a direct linkage between lattice models and fractal geometry, leading to innovative insights. The paper commences by elucidating the Ising model and a specialized classification scheme for its lattice points. Subsequently, it outlines the computational method for determining the weights associated with diverse lattice point categories, including those with spins oriented upwards and downwards. This enables the computation of thermodynamic phase transitions in both two-dimensional and three-dimensional Ising models. Moreover, the paper establishes a direct correlation between distinct types of lattice points and fundamental geometric principles, thereby revealing the intrinsic relationship between phase transitions and fractals. Finally, by fine-tuning the probabilistic computation method, the paper concludes that phase transitions and fractal structures emerge when the weights of boundary lattice points approach approximately 0.4.

\section{Theory}

\subsection{Ising model}

For each lattice site in the Ising model, the spin can exist in only two states: spin-up and spin-down. We can formulate the Hamiltonian for this system.

\begin{equation}
E(s_1,s_2,\dots,s_N) = \frac{1}{2}\sum_{i}^{N}\sum_{j}^{N} J_{ij} s_i s_j
\end{equation}
In the presence of an external magnetic field, this can be expressed as.
\begin{equation}
E(s_1,s_2,\dots,s_N) = \frac{1}{2}\sum_{i}^{N}\sum_{i}^{N} J_{ij} s_i s_j - H\sum_{i}^{N} s_i
\end{equation}
H represents magnetic field strength. We present the partition function.
\begin{equation}
Z = \sum_{s_1=\pm 1}\sum_{s_2=\pm 1}\dots\sum_{s_N=\pm 1} e^{\beta E(s_1,s_2,\dots,s_N)}
\end{equation}
The formula for the average magnetic field is given by
\begin{equation}
m=\frac{1}{N}<s_1+s_2+\dots+s_N>=<s_i>
\end{equation}

\subsection{Classification}
We adopt a classification approach based on the dissimilarity between each lattice point and its neighboring lattice points.
We categorize each lattice point based on the number of neighboring lattice points with the same spin as itself. Specifically, among the four nearest neighbor particles, if the number of particles with the same spin as the central lattice point is 1, it falls into a distinct class. Within this configuration, two possibilities exist: when the central spin is oriented upwards, one of the four neighboring lattice points has an upward spin while the remaining three have downward spins. Similarly, if the central lattice point's spin is downwards, one neighboring lattice point will have a downward spin while the other three will have upward spins. These two configurations belong to the same category. Following this rationale, we extend the categorization by considering the number of neighboring lattice points with the same spin as the central lattice point. By doing so, we divide all existing microscopic configurations into five classes. Each configuration possesses a distinct energy level. In the presence of an external magnetic field, the categorization extends to ten classes. In this context, we initially focus on scenarios without an external magnetic field. According to this classification, we can reformulate the Hamiltonian as follows.

\begin{eqnarray}
E(s_1,s_2,\dots,s_N)& = &\frac{1}{2}\sum_{i}^{m_1}\sum_{j}^{4} J_{ij} s_i s_j+\frac{1}{2}\sum_{i}^{m_2}\sum_{j}^{4} J_{ij} s_i s_j  \nonumber  \\
&\;&+ \dots+\frac{1}{2}\sum_{i}^{m_5}\sum_{j}^{4} J_{ij} s_i s_j  \nonumber  \\
\end{eqnarray}

\subsection{Detail balance}
In the preceding discussion, a classification scheme for lattice points in the Ising model was established based on the magnitudes of nearest-neighbor interactions. This categorization is subsequently linked to the phenomenon of phase transitions. For the sake of clarity, all lattice points were assumed to possess spins exclusively oriented either upwards or downwards at zero temperature. As the temperature rises, the ratio of lattice points with spins pointing downward (upward) is denoted as $m$.

Within the Ising model, interactions between adjacent lattice points are defined by a value of 1 if the spins are aligned and -1 if they are opposite, commonly referred to as positive and negative bonds, respectively. Throughout the temperature elevation process, the weights associated with positive and negative bonds undergo dynamic changes. The application of the detailed balance equation 
\begin{equation}
\sum_{\Gamma^{'}}\mathcal{T}(\Gamma^{'},\Gamma)P_{eq}(\Gamma) = \sum_{\Gamma}\mathcal{T}(\Gamma,\Gamma^{'})P_{eq}(\Gamma^{'})
\end{equation}

\begin{equation}
 \mathcal{T}(\Gamma^{'},\Gamma)P_{eq}(\Gamma) = \mathcal{T}(\Gamma,\Gamma^{'})P_{eq}(\Gamma^{'})  
 \label{Eq:2}
\end{equation}

enables the derivation of the weights of positive and negative bonds at different temperatures, offering valuable insights into the thermodynamic behavior of the Ising model.

In accounting for the impact of environmental factors during the transformation process of positive and negative bonds, the environmental influence in this process is approximated by $m$. The weights of positive and negative bonds are subsequently defined by Eq.~\ref{Eq:1}.

When a connecting rod associated with a lattice point is chosen, it results in $2d-1$ rods being connected to neighboring lattice points in its vicinity. The average magnetic field strength in the surrounding environment is approximated by $1-2m$.

\begin{equation}
 S=(2d-1)(1-2m)
 \label{Eq:1}
\end{equation}

\begin{equation}
k=\frac{1}{1+e^{-2S/T}}
\end{equation}

\begin{equation}
 c_i = C^{2d}_{i} k^{2d-i} (1-k)^{i-1}  
\end{equation}

The classification methodology elucidated earlier hinges on the enumeration of positive and negative bonds associated with each lattice point. Consequently, armed with the knowledge of the weights of positive and negative bonds, it becomes feasible to employ probability formulas for a direct computation of the weights pertaining to distinct categories of lattice points. This nuanced approach not only enhances our understanding of the intricate dynamics within the Ising model but also provides a systematic framework for evaluating the impact of environmental variables on bond transformations. These insights contribute to the broader landscape of statistical physics and pave the way for more sophisticated analyses in diverse scientific and technological domains.

When a lattice point and its nearest neighboring lattice point are selected, it results in $2d^2$ rods being connected to neighboring lattice points in its vicinity. The average magnetic field strength in the surrounding environment is approximated by $1-2m$.

\begin{equation}
 S1=2d^2(1-2m)  
\end{equation}

When the $\mathcal{T}$ is equal to or exceeds 1, the transformation is defined by 

\begin{equation}
 c_i\pm = c_i \times  \mathcal{T}(\Gamma^{'},\Gamma)=\frac{c_i}{1+e^{-2S1/T}}   
\end{equation}

 Conversely, when the $\mathcal{T}$ is less than 1, the transformation is expressed by 

\begin{equation}
 c_i\mp = c_i \times  \mathcal{T}(\Gamma^{'},\Gamma)=\frac{c_i\times e^{-2S1/T}}{1+e^{-2S1/T}}   
\end{equation}

When contemplating the polarities of each lattice point category, it is conceivable that positive and negative lattice points within a given category can mutually undergo transformations. By consistently incorporating the environmental influence approximated by $m$, the transitions between positive and negative lattice points adhere to the principles of detailed balance, as encapsulated by Eq.~\ref{Eq:2}

Consequently, the computation of the weights for diverse lattice point categories becomes feasible, enabling the determination of the weights associated with spins oriented upwards and downwards within each category. This intricate interconnection establishes a profound relationship between these five classifications and the observed phase transitions.

To corroborate the fidelity of this computational methodology with real-world scenarios, I applied this approach to calculate the thermodynamic phase transitions for both two-dimensional and three-dimensional Ising models as illustrated in Fig.~\ref{fig:2}. This rigorous examination not only validates the efficacy of the proposed method but also underscores its potential for advancing our understanding of the complex dynamics inherent in Ising models, offering a systematic framework for the analysis of phase transitions in various dimensions.

\begin{figure}
\centering
\begin{tikzpicture}

\scope[nodes={inner sep=4,outer sep=4}]
\node[anchor= east] (a)
  {\includegraphics[width=4cm]{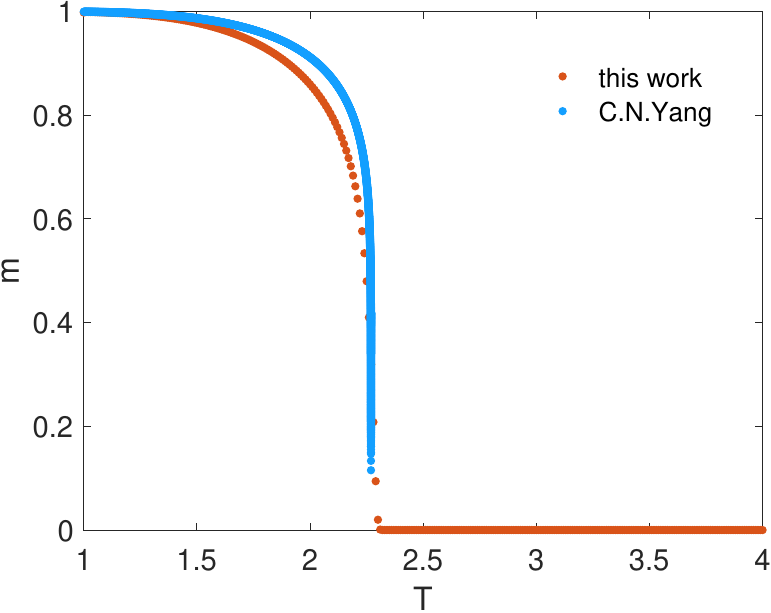}};
\node[anchor= west] (b)
  {\includegraphics[width=4cm]{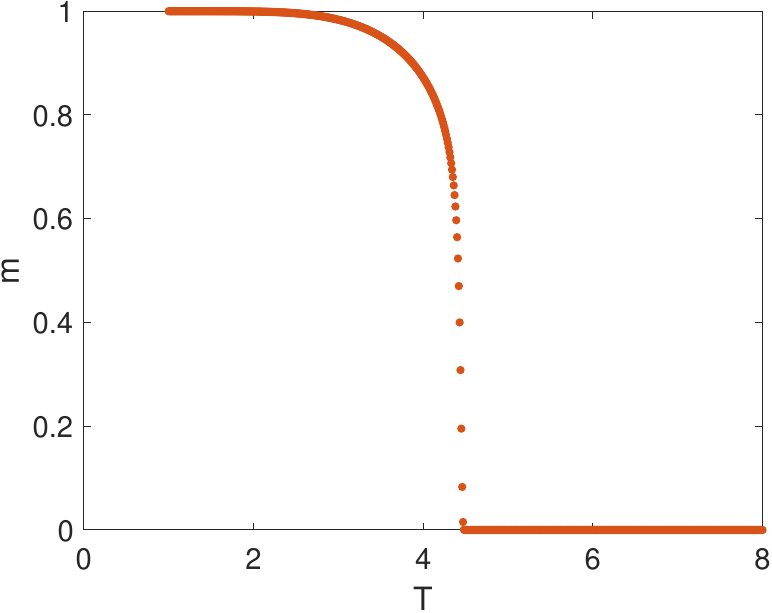}};
\endscope
\foreach \n in {a,b} {
  \node[anchor=west] at (\n.north west) {(\n)};
}

\end{tikzpicture}

\caption{\label{fig:2} a. The Temperature-Dependent Magnetic Susceptibility in the Two-Dimensional Ising Model. The blue dots represent the analytical solution of the two-dimensional Ising model, elucidated by C.N.Yang.
b. Temperature-Dependent Changes in Magnetic Susceptibility in the Three-Dimensional Ising Model. }
	\label{fig:2}
 
\end{figure}

\subsection{Fractal geometry}
Based on the computational findings presented earlier, it is evident that as the temperature rises to the phase transition point, a phase transition occurs, leading to each category of lattice points attaining specific weights. Concurrently, in both the two-dimensional and three-dimensional Ising models, the occurrence of phase transitions gives rise to fractal phenomena. More precisely, a fractal structure emerges when each category of lattice points reaches its designated weight. The subsequent section offers an illustrative exposition and elucidation of these relationships for enhanced conceptualization.

To delve into the intricate relationship between fractals and various lattice point categories, associations are established between differently classified lattice points and geometric entities. The distinguishing features of lattice points in the Ising model fall into three primary categories. The first category comprises lattice points entirely surrounded by spins of identical orientation. The second category encompasses boundary lattice points, where each point is proximate to only one lattice point with the opposite spin. In both two-dimensional and three-dimensional Ising models, the boundaries of sufficiently large regions with identical spins are predominantly constituted by lattice points falling into this category. The third category of lattice points encompasses all points not classified within the aforementioned two categories.

To provide specificity, in a two-dimensional Ising model, the first category corresponds to faces, the second category to edges, and the third category to structures involving lines and points within the faces. In a three-dimensional Ising model, the first category corresponds to volumes composed of spins with the same orientation, the second category corresponds to face lattice points on the boundaries of distinct volumes, and the third category corresponds to lines or points within the volumes as illustrated in Fig.~\ref{fig:3}. The weights of the second category of lattice points exhibit a unique behavior as the temperature increases, sharply decreasing upon reaching the phase transition point. This distinctive feature sets them apart from the other two categories of lattice points.

In summary, the Ising model's phase transition process manifests specific weights for these three categories of lattice points, each corresponding to distinct geometric structures. This multifaceted exploration not only advances our understanding of the Ising model's complex dynamics but also lays the groundwork for a systematic analysis of phase transitions across different dimensions, contributing to the broader landscape of statistical physics and applied research.

Upon surpassing the thermodynamic phase transition point, the system attains maximum entropy, thereby establishing the weights of different lattice points.

\begin{figure}
\centering
\begin{tikzpicture}

\scope[nodes={inner sep=4,outer sep=4}]
\node[anchor= east] (a)
  {\includegraphics[width=4cm]{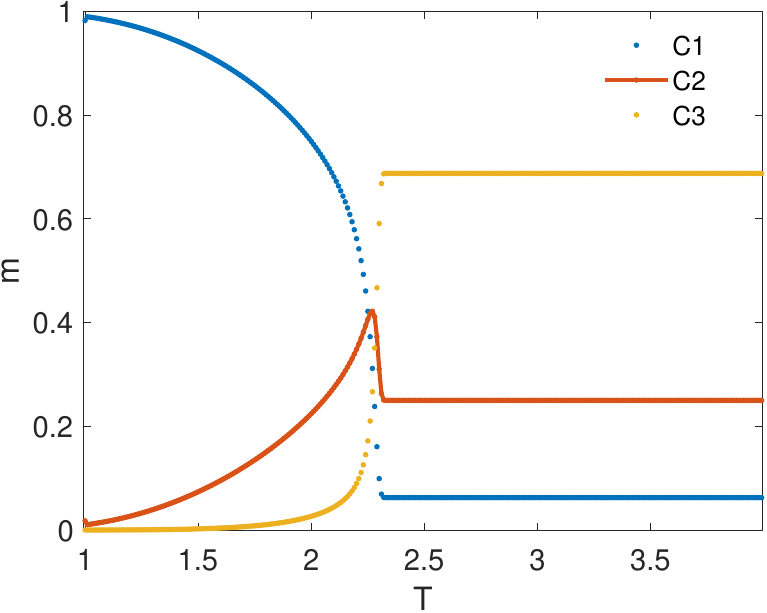}};
\node[anchor= west] (b)
  {\includegraphics[width=4cm]{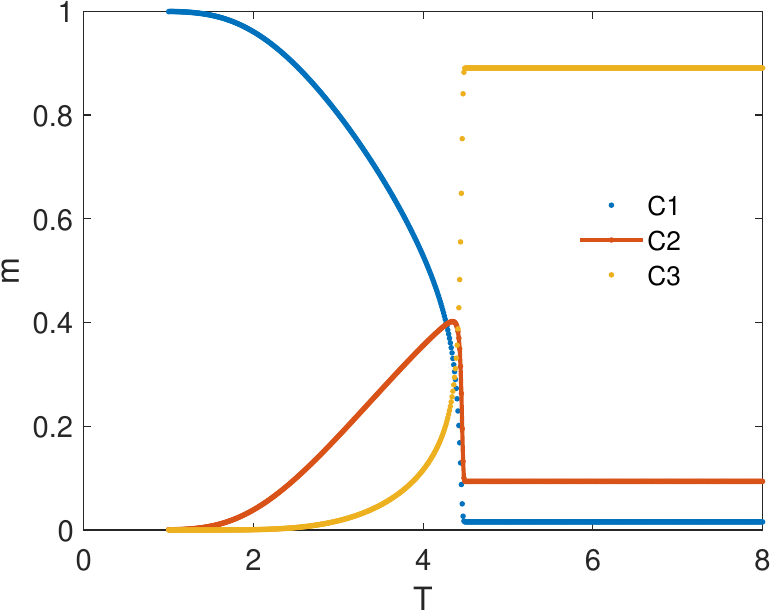}};
\endscope
\foreach \n in {a,b} {
  \node[anchor=west] at (\n.north west) {(\n)};
}

\end{tikzpicture}

\caption{\label{fig:3} a. The temperature-dependent evolution of weights for the three categories of lattice points in the two-dimensional Ising model reveals a distinctive behavior, notably observed in the peculiar phenomenon exhibited by the second category of lattice points.
b. In the three-dimensional Ising model, the weights of the three categories of lattice points exhibit temperature-dependent variations, and a notable phenomenon is observed in the peculiar behavior exhibited by the second category of lattice points. }
	\label{fig:3}
 
\end{figure}

\subsection{Transition point}
\begin{figure}
\centering
\includegraphics[width=0.7\linewidth]{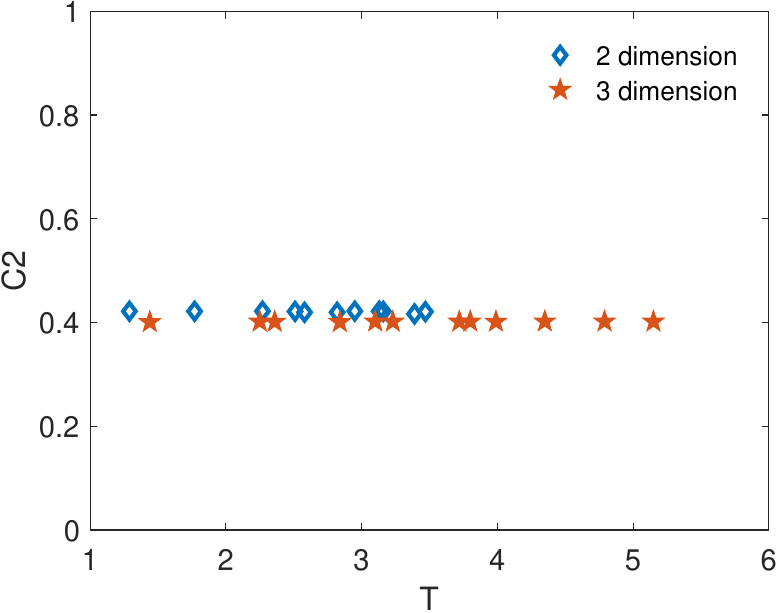}
\caption{\label{fig:4} The abscissa denotes the transition temperatures of thermodynamic phase transitions, whereas the ordinate represents the weights attributed to $C2$ at the phase transition points. The blue diamond markers correspond to the two-dimensional Ising model, while the orange pentagram markers correspond to the three-dimensional Ising model.}
\end{figure}

Drawing from the preceding results, we have acquired the weights corresponding to different categories of lattice points. To unravel the intricate relationships between these lattice point weights, phase transitions, and fractal phenomena, we systematically modulate the parameter $l$ to delineate the varying ratios of weights for different lattice point categories as temperature undergoes fluctuations. By introducing the parameter "$l$" to scale the values of $S$ and $S1$, we can effectively modulate the sizes of these parameters and, consequently, control the magnitude of the phase transition points. This suggests that points contributing to the formation of fractal structures undergo transformations akin to those occurring at the phase transition point. Such observations underscore a discernible interconnection between the two-dimensional and three-dimensional Ising models as illustrated in Fig.~\ref{fig:4}.

Furthermore, our investigations reveal that significant phase transitions and the manifestation of fractal structures occur prominently when boundary lattice points attain a specific threshold, establishing them as a pivotal determinant in fractal structure emergence. This nuanced exploration not only contributes to a comprehensive understanding of the dynamic interplay between lattice point weights, phase transitions, and fractal formations but also underscores the subtle yet crucial factors influencing such phenomena.

\section{Conclusion}
In the realm of the Ising model, grid points undergo classification based on the magnitude of their nearest-neighbor interactions, with the environmental influence approximated through magnetic induction strength. Leveraging the consideration of environmental impact, a meticulous balance equation is employed to ascertain the weights assigned to positive and negative rods. Utilizing a composite formula encapsulating both positive and negative rod weights, the weights for diverse grid point categories are computed. Furthermore, the spin orientation (up or down) for each grid point category can be elucidated via the detailed balance equation. As temperature ascends, probability formulas are presented, yielding expressions for the magnetic induction strength concerning two-dimensional and three-dimensional Ising models as a function of temperature. Impressively, results from the two-dimensional Ising model align notably well with analytical solutions, while the probability formula unveils phase transition points for the three-dimensional Ising model in close proximity to those obtained through Monte Carlo simulations.

The interconnection between distinct grid point categories and fundamental geometric concepts is systematically established, where each category corresponds to points, lines, planes, and other geometric entities. The categorization encompasses three classes of grid points, with the category housing grid points featuring precisely one neighboring spin with the opposite orientation denoted as boundary points. Illustrative graphs portraying the weight variations of the three categories of grid points in both two-dimensional and three-dimensional Ising models concerning temperature are provided. Evidently, all three categories exhibit abrupt changes at the phase transition point, with boundary points demonstrating an initial increase followed by a sudden decrease. Remarkably, owing to the maximum entropy principle, the weights of the three categories remain immutable post the phase transition point.

The introduction of a modulating factor ($l$) to adjust the size of the phase transition point is undertaken, showcasing that, for different values of $l$, the weights of boundary points in both two-dimensional and three-dimensional Ising models converge to approximately 0.4 at various phase transition points. Furthermore, it is deduced that around the 0.4 weight mark for boundary points in the Ising model, the emergence of fractal structures is attributed to the existence of phase transition points.

\begin{acknowledgments}
This paper is supported by the National Natural Science Foundation of China-China Academy of Engineering
Physics(CAEP)Joint Fund NSAF(No. U1930402).
\end{acknowledgments}

\nocite{*}
\bibliography{aipsamp}% Produces the bibliography via BibTeX.

\begin{thebibliography}{38}%
\makeatletter
\providecommand \@ifxundefined [1]{%
 \@ifx{#1\undefined}
}%
\providecommand \@ifnum [1]{%
 \ifnum #1\expandafter \@firstoftwo
 \else \expandafter \@secondoftwo
 \fi
}%
\providecommand \@ifx [1]{%
 \ifx #1\expandafter \@firstoftwo
 \else \expandafter \@secondoftwo
 \fi
}%
\providecommand \natexlab [1]{#1}%
\providecommand \enquote  [1]{``#1''}%
\providecommand \bibnamefont  [1]{#1}%
\providecommand \bibfnamefont [1]{#1}%
\providecommand \citenamefont [1]{#1}%
\providecommand \href@noop [0]{\@secondoftwo}%
\providecommand \href [0]{\begingroup \@sanitize@url \@href}%
\providecommand \@href[1]{\@@startlink{#1}\@@href}%
\providecommand \@@href[1]{\endgroup#1\@@endlink}%
\providecommand \@sanitize@url [0]{\catcode `\\12\catcode `\$12\catcode `\&12\catcode `\#12\catcode `\^12\catcode `\_12\catcode `\%12\relax}%
\providecommand \@@startlink[1]{}%
\providecommand \@@endlink[0]{}%
\providecommand \url  [0]{\begingroup\@sanitize@url \@url }%
\providecommand \@url [1]{\endgroup\@href {#1}{\urlprefix }}%
\providecommand \urlprefix  [0]{URL }%
\providecommand \Eprint [0]{\href }%
\providecommand \doibase [0]{http://dx.doi.org/}%
\providecommand \selectlanguage [0]{\@gobble}%
\providecommand \bibinfo  [0]{\@secondoftwo}%
\providecommand \bibfield  [0]{\@secondoftwo}%
\providecommand \translation [1]{[#1]}%
\providecommand \BibitemOpen [0]{}%
\providecommand \bibitemStop [0]{}%
\providecommand \bibitemNoStop [0]{.\EOS\space}%
\providecommand \EOS [0]{\spacefactor3000\relax}%
\providecommand \BibitemShut  [1]{\csname bibitem#1\endcsname}%
\let\auto@bib@innerbib\@empty
%</preamble>
\bibitem [{\citenamefont {Foulkes}\ \emph {et~al.}(2001)\citenamefont {Foulkes}, \citenamefont {Mitas}, \citenamefont {Needs},\ and\ \citenamefont {Rajagopal}}]{RevModPhys.73.33}%
  \BibitemOpen
  \bibfield  {author} {\bibinfo {author} {\bibfnamefont {W.~M.~C.}\ \bibnamefont {Foulkes}}, \bibinfo {author} {\bibfnamefont {L.}~\bibnamefont {Mitas}}, \bibinfo {author} {\bibfnamefont {R.~J.}\ \bibnamefont {Needs}}, \ and\ \bibinfo {author} {\bibfnamefont {G.}~\bibnamefont {Rajagopal}},\ }\bibfield  {title} {\enquote {\bibinfo {title} {Quantum monte carlo simulations of solids},}\ }\href {\doibase 10.1103/RevModPhys.73.33} {\bibfield  {journal} {\bibinfo  {journal} {Rev. Mod. Phys.}\ }\textbf {\bibinfo {volume} {73}},\ \bibinfo {pages} {33--83} (\bibinfo {year} {2001})}\BibitemShut {NoStop}%
\bibitem [{\citenamefont {Sarrut}\ \emph {et~al.}(2021)\citenamefont {Sarrut}, \citenamefont {Ba{\l}a}, \citenamefont {Bardi{\`e}s}, \citenamefont {Bert}, \citenamefont {Chauvin}, \citenamefont {Chatzipapas}, \citenamefont {Dupont}, \citenamefont {Etxebeste}, \citenamefont {Fanchon}, \citenamefont {Jan} \emph {et~al.}}]{sarrut2021advanced}%
  \BibitemOpen
  \bibfield  {author} {\bibinfo {author} {\bibfnamefont {D.}~\bibnamefont {Sarrut}}, \bibinfo {author} {\bibfnamefont {M.}~\bibnamefont {Ba{\l}a}}, \bibinfo {author} {\bibfnamefont {M.}~\bibnamefont {Bardi{\`e}s}}, \bibinfo {author} {\bibfnamefont {J.}~\bibnamefont {Bert}}, \bibinfo {author} {\bibfnamefont {M.}~\bibnamefont {Chauvin}}, \bibinfo {author} {\bibfnamefont {K.}~\bibnamefont {Chatzipapas}}, \bibinfo {author} {\bibfnamefont {M.}~\bibnamefont {Dupont}}, \bibinfo {author} {\bibfnamefont {A.}~\bibnamefont {Etxebeste}}, \bibinfo {author} {\bibfnamefont {L.~M.}\ \bibnamefont {Fanchon}}, \bibinfo {author} {\bibfnamefont {S.}~\bibnamefont {Jan}},  \emph {et~al.},\ }\bibfield  {title} {\enquote {\bibinfo {title} {Advanced monte carlo simulations of emission tomography imaging systems with gate},}\ }\href {\doibase 10.1088/1361-6560/abf276} {\bibfield  {journal} {\bibinfo  {journal} {Physics in Medicine \& Biology}\ }\textbf {\bibinfo {volume} {66}},\ \bibinfo {pages} {10TR03} (\bibinfo {year}
  {2021})}\BibitemShut {NoStop}%
\bibitem [{\citenamefont {Luo}\ \emph {et~al.}(2022)\citenamefont {Luo}, \citenamefont {Keshtegar}, \citenamefont {Zhu}, \citenamefont {Taylan},\ and\ \citenamefont {Niu}}]{luo2022hybrid}%
  \BibitemOpen
  \bibfield  {author} {\bibinfo {author} {\bibfnamefont {C.}~\bibnamefont {Luo}}, \bibinfo {author} {\bibfnamefont {B.}~\bibnamefont {Keshtegar}}, \bibinfo {author} {\bibfnamefont {S.~P.}\ \bibnamefont {Zhu}}, \bibinfo {author} {\bibfnamefont {O.}~\bibnamefont {Taylan}}, \ and\ \bibinfo {author} {\bibfnamefont {X.-P.}\ \bibnamefont {Niu}},\ }\bibfield  {title} {\enquote {\bibinfo {title} {Hybrid enhanced monte carlo simulation coupled with advanced machine learning approach for accurate and efficient structural reliability analysis},}\ }\href {\doibase https://doi.org/10.1016/j.cma.2021.114218} {\bibfield  {journal} {\bibinfo  {journal} {Computer Methods in Applied Mechanics and Engineering}\ }\textbf {\bibinfo {volume} {388}},\ \bibinfo {pages} {114218} (\bibinfo {year} {2022})}\BibitemShut {NoStop}%
\bibitem [{\citenamefont {Or{\'u}s}(2019)}]{orus2019tensor}%
  \BibitemOpen
  \bibfield  {author} {\bibinfo {author} {\bibfnamefont {R.}~\bibnamefont {Or{\'u}s}},\ }\bibfield  {title} {\enquote {\bibinfo {title} {Tensor networks for complex quantum systems},}\ }\href {\doibase https://ui.adsabs.harvard.edu/link_gateway/2019NatRP...1..538O/doi:10.1038/s42254-019-0086-7} {\bibfield  {journal} {\bibinfo  {journal} {Nature Reviews Physics}\ }\textbf {\bibinfo {volume} {1}},\ \bibinfo {pages} {538--550} (\bibinfo {year} {2019})}\BibitemShut {NoStop}%
\bibitem [{\citenamefont {Pan}\ and\ \citenamefont {Zhang}(2022)}]{pan2022simulation}%
  \BibitemOpen
  \bibfield  {author} {\bibinfo {author} {\bibfnamefont {F.}~\bibnamefont {Pan}}\ and\ \bibinfo {author} {\bibfnamefont {P.}~\bibnamefont {Zhang}},\ }\bibfield  {title} {\enquote {\bibinfo {title} {Simulation of quantum circuits using the big-batch tensor network method},}\ }\href {\doibase https://doi.org/10.1103/PhysRevLett.128.030501} {\bibfield  {journal} {\bibinfo  {journal} {Physical Review Letters}\ }\textbf {\bibinfo {volume} {128}},\ \bibinfo {pages} {030501} (\bibinfo {year} {2022})}\BibitemShut {NoStop}%
\bibitem [{\citenamefont {Gray}\ and\ \citenamefont {Kourtis}(2021)}]{gray2021hyper}%
  \BibitemOpen
  \bibfield  {author} {\bibinfo {author} {\bibfnamefont {J.}~\bibnamefont {Gray}}\ and\ \bibinfo {author} {\bibfnamefont {S.}~\bibnamefont {Kourtis}},\ }\bibfield  {title} {\enquote {\bibinfo {title} {Hyper-optimized tensor network contraction},}\ }\href {\doibase https://doi.org/10.22331/q-2021-03-15-410} {\bibfield  {journal} {\bibinfo  {journal} {Quantum}\ }\textbf {\bibinfo {volume} {5}},\ \bibinfo {pages} {410} (\bibinfo {year} {2021})}\BibitemShut {NoStop}%
\bibitem [{\citenamefont {Yuan}\ \emph {et~al.}(2021)\citenamefont {Yuan}, \citenamefont {Sun}, \citenamefont {Liu}, \citenamefont {Zhao},\ and\ \citenamefont {Zhou}}]{yuan2021quantum}%
  \BibitemOpen
  \bibfield  {author} {\bibinfo {author} {\bibfnamefont {X.}~\bibnamefont {Yuan}}, \bibinfo {author} {\bibfnamefont {J.}~\bibnamefont {Sun}}, \bibinfo {author} {\bibfnamefont {J.}~\bibnamefont {Liu}}, \bibinfo {author} {\bibfnamefont {Q.}~\bibnamefont {Zhao}}, \ and\ \bibinfo {author} {\bibfnamefont {Y.}~\bibnamefont {Zhou}},\ }\bibfield  {title} {\enquote {\bibinfo {title} {Quantum simulation with hybrid tensor networks},}\ }\href {\doibase https://doi.org/10.1103/PhysRevLett.127.040501} {\bibfield  {journal} {\bibinfo  {journal} {Physical Review Letters}\ }\textbf {\bibinfo {volume} {127}},\ \bibinfo {pages} {040501} (\bibinfo {year} {2021})}\BibitemShut {NoStop}%
\bibitem [{\citenamefont {Shao}, \citenamefont {Guo},\ and\ \citenamefont {Sandvik}(2016)}]{shao2016quantum}%
  \BibitemOpen
  \bibfield  {author} {\bibinfo {author} {\bibfnamefont {H.}~\bibnamefont {Shao}}, \bibinfo {author} {\bibfnamefont {W.}~\bibnamefont {Guo}}, \ and\ \bibinfo {author} {\bibfnamefont {A.~W.}\ \bibnamefont {Sandvik}},\ }\bibfield  {title} {\enquote {\bibinfo {title} {Quantum criticality with two length scales},}\ }\href {\doibase https://doi.org/10.1126/science.aad5007} {\bibfield  {journal} {\bibinfo  {journal} {Science}\ }\textbf {\bibinfo {volume} {352}},\ \bibinfo {pages} {213--216} (\bibinfo {year} {2016})}\BibitemShut {NoStop}%
\bibitem [{\citenamefont {Halperin}(2019)}]{halperin2019theory}%
  \BibitemOpen
  \bibfield  {author} {\bibinfo {author} {\bibfnamefont {B.~I.}\ \bibnamefont {Halperin}},\ }\bibfield  {title} {\enquote {\bibinfo {title} {Theory of dynamic critical phenomena},}\ }\href {\doibase https://doi.org/10.1063/PT.3.4137} {\bibfield  {journal} {\bibinfo  {journal} {Physics Today}\ }\textbf {\bibinfo {volume} {72}},\ \bibinfo {pages} {42--43} (\bibinfo {year} {2019})}\BibitemShut {NoStop}%
\bibitem [{\citenamefont {Manipatruni}\ \emph {et~al.}(2019)\citenamefont {Manipatruni}, \citenamefont {Nikonov}, \citenamefont {Lin}, \citenamefont {Gosavi}, \citenamefont {Liu}, \citenamefont {Prasad}, \citenamefont {Huang}, \citenamefont {Bonturim}, \citenamefont {Ramesh},\ and\ \citenamefont {Young}}]{manipatruni2019scalable}%
  \BibitemOpen
  \bibfield  {author} {\bibinfo {author} {\bibfnamefont {S.}~\bibnamefont {Manipatruni}}, \bibinfo {author} {\bibfnamefont {D.~E.}\ \bibnamefont {Nikonov}}, \bibinfo {author} {\bibfnamefont {C.-C.}\ \bibnamefont {Lin}}, \bibinfo {author} {\bibfnamefont {T.~A.}\ \bibnamefont {Gosavi}}, \bibinfo {author} {\bibfnamefont {H.}~\bibnamefont {Liu}}, \bibinfo {author} {\bibfnamefont {B.}~\bibnamefont {Prasad}}, \bibinfo {author} {\bibfnamefont {Y.-L.}\ \bibnamefont {Huang}}, \bibinfo {author} {\bibfnamefont {E.}~\bibnamefont {Bonturim}}, \bibinfo {author} {\bibfnamefont {R.}~\bibnamefont {Ramesh}}, \ and\ \bibinfo {author} {\bibfnamefont {I.~A.}\ \bibnamefont {Young}},\ }\bibfield  {title} {\enquote {\bibinfo {title} {Scalable energy-efficient magnetoelectric spin--orbit logic},}\ }\href {\doibase https://doi.org/10.1038/s41586-018-0770-2} {\bibfield  {journal} {\bibinfo  {journal} {Nature}\ }\textbf {\bibinfo {volume} {565}},\ \bibinfo {pages} {35--42} (\bibinfo {year} {2019})}\BibitemShut {NoStop}%
\bibitem [{\citenamefont {Tang}\ \emph {et~al.}(2018)\citenamefont {Tang}, \citenamefont {Leaw}, \citenamefont {Rodrigues}, \citenamefont {Herbut}, \citenamefont {Sengupta}, \citenamefont {Assaad},\ and\ \citenamefont {Adam}}]{tang2018role}%
  \BibitemOpen
  \bibfield  {author} {\bibinfo {author} {\bibfnamefont {H.-K.}\ \bibnamefont {Tang}}, \bibinfo {author} {\bibfnamefont {J.}~\bibnamefont {Leaw}}, \bibinfo {author} {\bibfnamefont {J.}~\bibnamefont {Rodrigues}}, \bibinfo {author} {\bibfnamefont {I.}~\bibnamefont {Herbut}}, \bibinfo {author} {\bibfnamefont {P.}~\bibnamefont {Sengupta}}, \bibinfo {author} {\bibfnamefont {F.}~\bibnamefont {Assaad}}, \ and\ \bibinfo {author} {\bibfnamefont {S.}~\bibnamefont {Adam}},\ }\bibfield  {title} {\enquote {\bibinfo {title} {The role of electron-electron interactions in two-dimensional dirac fermions},}\ }\href {\doibase https://doi.org/10.1126/science.aao2934} {\bibfield  {journal} {\bibinfo  {journal} {Science}\ }\textbf {\bibinfo {volume} {361}},\ \bibinfo {pages} {570--574} (\bibinfo {year} {2018})}\BibitemShut {NoStop}%
\bibitem [{\citenamefont {Carlson}\ \emph {et~al.}(2015)\citenamefont {Carlson}, \citenamefont {Gandolfi}, \citenamefont {Pederiva}, \citenamefont {Pieper}, \citenamefont {Schiavilla}, \citenamefont {Schmidt},\ and\ \citenamefont {Wiringa}}]{carlson2015quantum}%
  \BibitemOpen
  \bibfield  {author} {\bibinfo {author} {\bibfnamefont {J.}~\bibnamefont {Carlson}}, \bibinfo {author} {\bibfnamefont {S.}~\bibnamefont {Gandolfi}}, \bibinfo {author} {\bibfnamefont {F.}~\bibnamefont {Pederiva}}, \bibinfo {author} {\bibfnamefont {S.~C.}\ \bibnamefont {Pieper}}, \bibinfo {author} {\bibfnamefont {R.}~\bibnamefont {Schiavilla}}, \bibinfo {author} {\bibfnamefont {K.~E.}\ \bibnamefont {Schmidt}}, \ and\ \bibinfo {author} {\bibfnamefont {R.~B.}\ \bibnamefont {Wiringa}},\ }\bibfield  {title} {\enquote {\bibinfo {title} {Quantum monte carlo methods for nuclear physics},}\ }\href {\doibase https://doi.org/10.1103/RevModPhys.87.1067} {\bibfield  {journal} {\bibinfo  {journal} {Reviews of Modern Physics}\ }\textbf {\bibinfo {volume} {87}},\ \bibinfo {pages} {1067} (\bibinfo {year} {2015})}\BibitemShut {NoStop}%
\bibitem [{\citenamefont {Lynn}\ \emph {et~al.}(2019)\citenamefont {Lynn}, \citenamefont {Tews}, \citenamefont {Gandolfi},\ and\ \citenamefont {Lovato}}]{lynn2019quantum}%
  \BibitemOpen
  \bibfield  {author} {\bibinfo {author} {\bibfnamefont {J.~E.}\ \bibnamefont {Lynn}}, \bibinfo {author} {\bibfnamefont {I.}~\bibnamefont {Tews}}, \bibinfo {author} {\bibfnamefont {S.}~\bibnamefont {Gandolfi}}, \ and\ \bibinfo {author} {\bibfnamefont {A.}~\bibnamefont {Lovato}},\ }\bibfield  {title} {\enquote {\bibinfo {title} {Quantum monte carlo methods in nuclear physics: recent advances},}\ }\href {\doibase https://doi.org/10.1146/annurev-nucl-101918-023600} {\bibfield  {journal} {\bibinfo  {journal} {Annual Review of Nuclear and Particle Science}\ }\textbf {\bibinfo {volume} {69}},\ \bibinfo {pages} {279--305} (\bibinfo {year} {2019})}\BibitemShut {NoStop}%
\bibitem [{\citenamefont {Nemeth}\ and\ \citenamefont {Fearnhead}(2021)}]{nemeth2021stochastic}%
  \BibitemOpen
  \bibfield  {author} {\bibinfo {author} {\bibfnamefont {C.}~\bibnamefont {Nemeth}}\ and\ \bibinfo {author} {\bibfnamefont {P.}~\bibnamefont {Fearnhead}},\ }\bibfield  {title} {\enquote {\bibinfo {title} {Stochastic gradient markov chain monte carlo},}\ }\href {\doibase https://doi.org/10.1080/01621459.2020.1847120} {\bibfield  {journal} {\bibinfo  {journal} {Journal of the American Statistical Association}\ }\textbf {\bibinfo {volume} {116}},\ \bibinfo {pages} {433--450} (\bibinfo {year} {2021})}\BibitemShut {NoStop}%
\bibitem [{\citenamefont {Rioux-Lavoie}\ \emph {et~al.}(2022)\citenamefont {Rioux-Lavoie}, \citenamefont {Sugimoto}, \citenamefont {{\"O}zdemir}, \citenamefont {Shimada}, \citenamefont {Batty}, \citenamefont {Nowrouzezahrai},\ and\ \citenamefont {Hachisuka}}]{rioux2022monte}%
  \BibitemOpen
  \bibfield  {author} {\bibinfo {author} {\bibfnamefont {D.}~\bibnamefont {Rioux-Lavoie}}, \bibinfo {author} {\bibfnamefont {R.}~\bibnamefont {Sugimoto}}, \bibinfo {author} {\bibfnamefont {T.}~\bibnamefont {{\"O}zdemir}}, \bibinfo {author} {\bibfnamefont {N.~H.}\ \bibnamefont {Shimada}}, \bibinfo {author} {\bibfnamefont {C.}~\bibnamefont {Batty}}, \bibinfo {author} {\bibfnamefont {D.}~\bibnamefont {Nowrouzezahrai}}, \ and\ \bibinfo {author} {\bibfnamefont {T.}~\bibnamefont {Hachisuka}},\ }\bibfield  {title} {\enquote {\bibinfo {title} {A monte carlo method for fluid simulation},}\ }\href {\doibase https://doi.org/10.1145/3550454.3555450} {\bibfield  {journal} {\bibinfo  {journal} {ACM Transactions on Graphics (TOG)}\ }\textbf {\bibinfo {volume} {41}},\ \bibinfo {pages} {1--16} (\bibinfo {year} {2022})}\BibitemShut {NoStop}%
\bibitem [{\citenamefont {Andersen}, \citenamefont {Panosetti},\ and\ \citenamefont {Reuter}(2019)}]{andersen2019practical}%
  \BibitemOpen
  \bibfield  {author} {\bibinfo {author} {\bibfnamefont {M.}~\bibnamefont {Andersen}}, \bibinfo {author} {\bibfnamefont {C.}~\bibnamefont {Panosetti}}, \ and\ \bibinfo {author} {\bibfnamefont {K.}~\bibnamefont {Reuter}},\ }\bibfield  {title} {\enquote {\bibinfo {title} {A practical guide to surface kinetic monte carlo simulations},}\ }\href {\doibase https://doi.org/10.3389/fchem.2019.00202} {\bibfield  {journal} {\bibinfo  {journal} {Frontiers in chemistry}\ }\textbf {\bibinfo {volume} {7}},\ \bibinfo {pages} {202} (\bibinfo {year} {2019})}\BibitemShut {NoStop}%
\bibitem [{\citenamefont {Barouch}, \citenamefont {McCoy},\ and\ \citenamefont {Dresden}(1970)}]{barouch1970statistical}%
  \BibitemOpen
  \bibfield  {author} {\bibinfo {author} {\bibfnamefont {E.}~\bibnamefont {Barouch}}, \bibinfo {author} {\bibfnamefont {B.~M.}\ \bibnamefont {McCoy}}, \ and\ \bibinfo {author} {\bibfnamefont {M.}~\bibnamefont {Dresden}},\ }\bibfield  {title} {\enquote {\bibinfo {title} {Statistical mechanics of the xy model. i},}\ }\href {\doibase https://doi.org/10.1103/PhysRevA.2.1075} {\bibfield  {journal} {\bibinfo  {journal} {Physical Review A}\ }\textbf {\bibinfo {volume} {2}},\ \bibinfo {pages} {1075} (\bibinfo {year} {1970})}\BibitemShut {NoStop}%
\bibitem [{\citenamefont {Kosterlitz}(1974)}]{kosterlitz1974critical}%
  \BibitemOpen
  \bibfield  {author} {\bibinfo {author} {\bibfnamefont {J.}~\bibnamefont {Kosterlitz}},\ }\bibfield  {title} {\enquote {\bibinfo {title} {The critical properties of the two-dimensional xy model},}\ }\href {\doibase 10.1088/0022-3719/7/6/005} {\bibfield  {journal} {\bibinfo  {journal} {Journal of Physics C: Solid State Physics}\ }\textbf {\bibinfo {volume} {7}},\ \bibinfo {pages} {1046} (\bibinfo {year} {1974})}\BibitemShut {NoStop}%
\bibitem [{\citenamefont {Desai}\ and\ \citenamefont {Kaul}(2020)}]{desai2020first}%
  \BibitemOpen
  \bibfield  {author} {\bibinfo {author} {\bibfnamefont {N.}~\bibnamefont {Desai}}\ and\ \bibinfo {author} {\bibfnamefont {R.~K.}\ \bibnamefont {Kaul}},\ }\bibfield  {title} {\enquote {\bibinfo {title} {First-order phase transitions in the square-lattice easy-plane jq model},}\ }\href {\doibase https://doi.org/10.1103/PhysRevB.102.195135} {\bibfield  {journal} {\bibinfo  {journal} {Physical Review B}\ }\textbf {\bibinfo {volume} {102}},\ \bibinfo {pages} {195135} (\bibinfo {year} {2020})}\BibitemShut {NoStop}%
\bibitem [{\citenamefont {Wu}(1982)}]{wu1982potts}%
  \BibitemOpen
  \bibfield  {author} {\bibinfo {author} {\bibfnamefont {F.-Y.}\ \bibnamefont {Wu}},\ }\bibfield  {title} {\enquote {\bibinfo {title} {The potts model},}\ }\href {\doibase https://doi.org/10.1103/RevModPhys.54.235} {\bibfield  {journal} {\bibinfo  {journal} {Reviews of modern physics}\ }\textbf {\bibinfo {volume} {54}},\ \bibinfo {pages} {235} (\bibinfo {year} {1982})}\BibitemShut {NoStop}%
\bibitem [{\citenamefont {Henelius}\ and\ \citenamefont {Sandvik}(2000)}]{henelius2000sign}%
  \BibitemOpen
  \bibfield  {author} {\bibinfo {author} {\bibfnamefont {P.}~\bibnamefont {Henelius}}\ and\ \bibinfo {author} {\bibfnamefont {A.~W.}\ \bibnamefont {Sandvik}},\ }\bibfield  {title} {\enquote {\bibinfo {title} {Sign problem in monte carlo simulations of frustrated quantum spin systems},}\ }\href {\doibase https://doi.org/10.1103/PhysRevB.62.1102} {\bibfield  {journal} {\bibinfo  {journal} {Physical Review B}\ }\textbf {\bibinfo {volume} {62}},\ \bibinfo {pages} {1102} (\bibinfo {year} {2000})}\BibitemShut {NoStop}%
\bibitem [{\citenamefont {Alexandru}\ \emph {et~al.}(2022)\citenamefont {Alexandru}, \citenamefont {Ba{\c{s}}ar}, \citenamefont {Bedaque},\ and\ \citenamefont {Warrington}}]{alexandru2022complex}%
  \BibitemOpen
  \bibfield  {author} {\bibinfo {author} {\bibfnamefont {A.}~\bibnamefont {Alexandru}}, \bibinfo {author} {\bibfnamefont {G.}~\bibnamefont {Ba{\c{s}}ar}}, \bibinfo {author} {\bibfnamefont {P.~F.}\ \bibnamefont {Bedaque}}, \ and\ \bibinfo {author} {\bibfnamefont {N.~C.}\ \bibnamefont {Warrington}},\ }\bibfield  {title} {\enquote {\bibinfo {title} {Complex paths around the sign problem},}\ }\href {\doibase https://doi.org/10.1103/RevModPhys.94.015006} {\bibfield  {journal} {\bibinfo  {journal} {Reviews of Modern Physics}\ }\textbf {\bibinfo {volume} {94}},\ \bibinfo {pages} {015006} (\bibinfo {year} {2022})}\BibitemShut {NoStop}%
\bibitem [{\citenamefont {Mondaini}, \citenamefont {Tarat},\ and\ \citenamefont {Scalettar}(2022)}]{mondaini2022quantum}%
  \BibitemOpen
  \bibfield  {author} {\bibinfo {author} {\bibfnamefont {R.}~\bibnamefont {Mondaini}}, \bibinfo {author} {\bibfnamefont {S.}~\bibnamefont {Tarat}}, \ and\ \bibinfo {author} {\bibfnamefont {R.~T.}\ \bibnamefont {Scalettar}},\ }\bibfield  {title} {\enquote {\bibinfo {title} {Quantum critical points and the sign problem},}\ }\href {\doibase https://doi.org/10.1126/science.abg9299} {\bibfield  {journal} {\bibinfo  {journal} {Science}\ }\textbf {\bibinfo {volume} {375}},\ \bibinfo {pages} {418--424} (\bibinfo {year} {2022})}\BibitemShut {NoStop}%
\bibitem [{\citenamefont {Zhang}\ \emph {et~al.}(2022)\citenamefont {Zhang}, \citenamefont {Pan}, \citenamefont {Xu},\ and\ \citenamefont {Meng}}]{zhang2022fermion}%
  \BibitemOpen
  \bibfield  {author} {\bibinfo {author} {\bibfnamefont {X.}~\bibnamefont {Zhang}}, \bibinfo {author} {\bibfnamefont {G.}~\bibnamefont {Pan}}, \bibinfo {author} {\bibfnamefont {X.~Y.}\ \bibnamefont {Xu}}, \ and\ \bibinfo {author} {\bibfnamefont {Z.~Y.}\ \bibnamefont {Meng}},\ }\bibfield  {title} {\enquote {\bibinfo {title} {Fermion sign bounds theory in quantum monte carlo simulation},}\ }\href {\doibase https://doi.org/10.1103/PhysRevB.106.035121} {\bibfield  {journal} {\bibinfo  {journal} {Physical Review B}\ }\textbf {\bibinfo {volume} {106}},\ \bibinfo {pages} {035121} (\bibinfo {year} {2022})}\BibitemShut {NoStop}%
\bibitem [{\citenamefont {Berger}\ \emph {et~al.}(2021)\citenamefont {Berger}, \citenamefont {Rammelm{\"u}ller}, \citenamefont {Loheac}, \citenamefont {Ehmann}, \citenamefont {Braun},\ and\ \citenamefont {Drut}}]{berger2021complex}%
  \BibitemOpen
  \bibfield  {author} {\bibinfo {author} {\bibfnamefont {C.~E.}\ \bibnamefont {Berger}}, \bibinfo {author} {\bibfnamefont {L.}~\bibnamefont {Rammelm{\"u}ller}}, \bibinfo {author} {\bibfnamefont {A.~C.}\ \bibnamefont {Loheac}}, \bibinfo {author} {\bibfnamefont {F.}~\bibnamefont {Ehmann}}, \bibinfo {author} {\bibfnamefont {J.}~\bibnamefont {Braun}}, \ and\ \bibinfo {author} {\bibfnamefont {J.~E.}\ \bibnamefont {Drut}},\ }\bibfield  {title} {\enquote {\bibinfo {title} {Complex langevin and other approaches to the sign problem in quantum many-body physics},}\ }\href {\doibase https://doi.org/10.1016/j.physrep.2020.09.002} {\bibfield  {journal} {\bibinfo  {journal} {Physics Reports}\ }\textbf {\bibinfo {volume} {892}},\ \bibinfo {pages} {1--54} (\bibinfo {year} {2021})}\BibitemShut {NoStop}%
\bibitem [{\citenamefont {Kong}\ \emph {et~al.}(2020)\citenamefont {Kong}, \citenamefont {Lan}, \citenamefont {Wen}, \citenamefont {Zhang},\ and\ \citenamefont {Zheng}}]{kong2020algebraic}%
  \BibitemOpen
  \bibfield  {author} {\bibinfo {author} {\bibfnamefont {L.}~\bibnamefont {Kong}}, \bibinfo {author} {\bibfnamefont {T.}~\bibnamefont {Lan}}, \bibinfo {author} {\bibfnamefont {X.-G.}\ \bibnamefont {Wen}}, \bibinfo {author} {\bibfnamefont {Z.-H.}\ \bibnamefont {Zhang}}, \ and\ \bibinfo {author} {\bibfnamefont {H.}~\bibnamefont {Zheng}},\ }\bibfield  {title} {\enquote {\bibinfo {title} {Algebraic higher symmetry and categorical symmetry: A holographic and entanglement view of symmetry},}\ }\href {\doibase https://doi.org/10.1103/PhysRevResearch.2.043086} {\bibfield  {journal} {\bibinfo  {journal} {Physical Review Research}\ }\textbf {\bibinfo {volume} {2}},\ \bibinfo {pages} {043086} (\bibinfo {year} {2020})}\BibitemShut {NoStop}%
\bibitem [{\citenamefont {Hu}\ \emph {et~al.}(2020)\citenamefont {Hu}, \citenamefont {Wang}, \citenamefont {Zhao},\ and\ \citenamefont {Deng}}]{hu2020symmetry}%
  \BibitemOpen
  \bibfield  {author} {\bibinfo {author} {\bibfnamefont {W.}~\bibnamefont {Hu}}, \bibinfo {author} {\bibfnamefont {Z.}~\bibnamefont {Wang}}, \bibinfo {author} {\bibfnamefont {Y.}~\bibnamefont {Zhao}}, \ and\ \bibinfo {author} {\bibfnamefont {Z.}~\bibnamefont {Deng}},\ }\bibfield  {title} {\enquote {\bibinfo {title} {Symmetry breaking of infinite-dimensional dynamic system},}\ }\href {\doibase https://doi.org/10.1016/j.aml.2019.106207} {\bibfield  {journal} {\bibinfo  {journal} {Applied Mathematics Letters}\ }\textbf {\bibinfo {volume} {103}},\ \bibinfo {pages} {106207} (\bibinfo {year} {2020})}\BibitemShut {NoStop}%
\bibitem [{\citenamefont {Ortega-Zamorano}\ \emph {et~al.}(2015)\citenamefont {Ortega-Zamorano}, \citenamefont {Montemurro}, \citenamefont {Cannas}, \citenamefont {Jerez},\ and\ \citenamefont {Franco}}]{ortega2015fpga}%
  \BibitemOpen
  \bibfield  {author} {\bibinfo {author} {\bibfnamefont {F.}~\bibnamefont {Ortega-Zamorano}}, \bibinfo {author} {\bibfnamefont {M.~A.}\ \bibnamefont {Montemurro}}, \bibinfo {author} {\bibfnamefont {S.~A.}\ \bibnamefont {Cannas}}, \bibinfo {author} {\bibfnamefont {J.~M.}\ \bibnamefont {Jerez}}, \ and\ \bibinfo {author} {\bibfnamefont {L.}~\bibnamefont {Franco}},\ }\bibfield  {title} {\enquote {\bibinfo {title} {Fpga hardware acceleration of monte carlo simulations for the ising model},}\ }\href {\doibase https://doi.org/10.1109/TPDS.2015.2505725} {\bibfield  {journal} {\bibinfo  {journal} {IEEE Transactions on Parallel and Distributed Systems}\ }\textbf {\bibinfo {volume} {27}},\ \bibinfo {pages} {2618--2627} (\bibinfo {year} {2015})}\BibitemShut {NoStop}%
\bibitem [{\citenamefont {Yang}\ \emph {et~al.}(2019)\citenamefont {Yang}, \citenamefont {Chen}, \citenamefont {Roumpos}, \citenamefont {Colby},\ and\ \citenamefont {Anderson}}]{yang2019high}%
  \BibitemOpen
  \bibfield  {author} {\bibinfo {author} {\bibfnamefont {K.}~\bibnamefont {Yang}}, \bibinfo {author} {\bibfnamefont {Y.-F.}\ \bibnamefont {Chen}}, \bibinfo {author} {\bibfnamefont {G.}~\bibnamefont {Roumpos}}, \bibinfo {author} {\bibfnamefont {C.}~\bibnamefont {Colby}}, \ and\ \bibinfo {author} {\bibfnamefont {J.}~\bibnamefont {Anderson}},\ }\bibfield  {title} {\enquote {\bibinfo {title} {High performance monte carlo simulation of ising model on tpu clusters},}\ }in\ \href {\doibase https://doi.org/10.1145/3295500.3356149} {\emph {\bibinfo {booktitle} {Proceedings of the International Conference for High Performance Computing, Networking, Storage and Analysis}}}\ (\bibinfo {year} {2019})\ pp.\ \bibinfo {pages} {1--15}\BibitemShut {NoStop}%
\bibitem [{\citenamefont {Preis}\ \emph {et~al.}(2009)\citenamefont {Preis}, \citenamefont {Virnau}, \citenamefont {Paul},\ and\ \citenamefont {Schneider}}]{preis2009gpu}%
  \BibitemOpen
  \bibfield  {author} {\bibinfo {author} {\bibfnamefont {T.}~\bibnamefont {Preis}}, \bibinfo {author} {\bibfnamefont {P.}~\bibnamefont {Virnau}}, \bibinfo {author} {\bibfnamefont {W.}~\bibnamefont {Paul}}, \ and\ \bibinfo {author} {\bibfnamefont {J.~J.}\ \bibnamefont {Schneider}},\ }\bibfield  {title} {\enquote {\bibinfo {title} {Gpu accelerated monte carlo simulation of the 2d and 3d ising model},}\ }\href {\doibase https://doi.org/10.1016/j.jcp.2009.03.018} {\bibfield  {journal} {\bibinfo  {journal} {Journal of Computational Physics}\ }\textbf {\bibinfo {volume} {228}},\ \bibinfo {pages} {4468--4477} (\bibinfo {year} {2009})}\BibitemShut {NoStop}%
\bibitem [{\citenamefont {Suzuki}, \citenamefont {Nishimori},\ and\ \citenamefont {Suzuki}(2007)}]{suzuki2007quantum}%
  \BibitemOpen
  \bibfield  {author} {\bibinfo {author} {\bibfnamefont {S.}~\bibnamefont {Suzuki}}, \bibinfo {author} {\bibfnamefont {H.}~\bibnamefont {Nishimori}}, \ and\ \bibinfo {author} {\bibfnamefont {M.}~\bibnamefont {Suzuki}},\ }\bibfield  {title} {\enquote {\bibinfo {title} {Quantum annealing of the random-field ising model by transverse ferromagnetic interactions},}\ }\href {\doibase https://doi.org/10.1103/PhysRevE.75.051112} {\bibfield  {journal} {\bibinfo  {journal} {Physical Review E}\ }\textbf {\bibinfo {volume} {75}},\ \bibinfo {pages} {051112} (\bibinfo {year} {2007})}\BibitemShut {NoStop}%
\bibitem [{\citenamefont {Dziarmaga}(2005)}]{dziarmaga2005dynamics}%
  \BibitemOpen
  \bibfield  {author} {\bibinfo {author} {\bibfnamefont {J.}~\bibnamefont {Dziarmaga}},\ }\bibfield  {title} {\enquote {\bibinfo {title} {Dynamics of a quantum phase transition: Exact solution of the quantum ising model},}\ }\href {\doibase https://doi.org/10.1103/PhysRevLett.95.245701} {\bibfield  {journal} {\bibinfo  {journal} {Physical review letters}\ }\textbf {\bibinfo {volume} {95}},\ \bibinfo {pages} {245701} (\bibinfo {year} {2005})}\BibitemShut {NoStop}%
\bibitem [{\citenamefont {Baxter}\ and\ \citenamefont {Enting}(1978)}]{baxter1978399th}%
  \BibitemOpen
  \bibfield  {author} {\bibinfo {author} {\bibfnamefont {R.~J.}\ \bibnamefont {Baxter}}\ and\ \bibinfo {author} {\bibfnamefont {I.~G.}\ \bibnamefont {Enting}},\ }\bibfield  {title} {\enquote {\bibinfo {title} {399th solution of the ising model},}\ }\href {\doibase 10.1088/0305-4470/11/12/012} {\bibfield  {journal} {\bibinfo  {journal} {Journal of Physics A: Mathematical and General}\ }\textbf {\bibinfo {volume} {11}},\ \bibinfo {pages} {2463} (\bibinfo {year} {1978})}\BibitemShut {NoStop}%
\bibitem [{\citenamefont {Zhang}\ \emph {et~al.}(2020)\citenamefont {Zhang}, \citenamefont {Michel}, \citenamefont {El{\c{c}}i},\ and\ \citenamefont {Deng}}]{zhang2020loop}%
  \BibitemOpen
  \bibfield  {author} {\bibinfo {author} {\bibfnamefont {L.}~\bibnamefont {Zhang}}, \bibinfo {author} {\bibfnamefont {M.}~\bibnamefont {Michel}}, \bibinfo {author} {\bibfnamefont {E.~M.}\ \bibnamefont {El{\c{c}}i}}, \ and\ \bibinfo {author} {\bibfnamefont {Y.}~\bibnamefont {Deng}},\ }\bibfield  {title} {\enquote {\bibinfo {title} {Loop-cluster coupling and algorithm for classical statistical models},}\ }\href {\doibase https://doi.org/10.1103/PhysRevLett.125.200603} {\bibfield  {journal} {\bibinfo  {journal} {Physical Review Letters}\ }\textbf {\bibinfo {volume} {125}},\ \bibinfo {pages} {200603} (\bibinfo {year} {2020})}\BibitemShut {NoStop}%
\bibitem [{\citenamefont {Fang}, \citenamefont {Zhou},\ and\ \citenamefont {Deng}(2023)}]{fang2023geometric}%
  \BibitemOpen
  \bibfield  {author} {\bibinfo {author} {\bibfnamefont {S.}~\bibnamefont {Fang}}, \bibinfo {author} {\bibfnamefont {Z.}~\bibnamefont {Zhou}}, \ and\ \bibinfo {author} {\bibfnamefont {Y.}~\bibnamefont {Deng}},\ }\bibfield  {title} {\enquote {\bibinfo {title} {Geometric scaling behaviors of the fortuin-kasteleyn ising model in high dimensions},}\ }\href {\doibase https://doi.org/10.1103/PhysRevE.107.044103} {\bibfield  {journal} {\bibinfo  {journal} {Physical Review E}\ }\textbf {\bibinfo {volume} {107}},\ \bibinfo {pages} {044103} (\bibinfo {year} {2023})}\BibitemShut {NoStop}%
\bibitem [{\citenamefont {Park}, \citenamefont {Jin},\ and\ \citenamefont {Schweinberger}(2022)}]{park2022bayesian}%
  \BibitemOpen
  \bibfield  {author} {\bibinfo {author} {\bibfnamefont {J.}~\bibnamefont {Park}}, \bibinfo {author} {\bibfnamefont {I.~H.}\ \bibnamefont {Jin}}, \ and\ \bibinfo {author} {\bibfnamefont {M.}~\bibnamefont {Schweinberger}},\ }\bibfield  {title} {\enquote {\bibinfo {title} {Bayesian model selection for high-dimensional ising models, with applications to educational data},}\ }\href {\doibase https://doi.org/10.1016/j.csda.2021.107325} {\bibfield  {journal} {\bibinfo  {journal} {Computational Statistics \& Data Analysis}\ }\textbf {\bibinfo {volume} {165}},\ \bibinfo {pages} {107325} (\bibinfo {year} {2022})}\BibitemShut {NoStop}%
\bibitem [{\citenamefont {Kryzhanovsky}, \citenamefont {Litinskii},\ and\ \citenamefont {Egorov}(2021)}]{kryzhanovsky2021analytical}%
  \BibitemOpen
  \bibfield  {author} {\bibinfo {author} {\bibfnamefont {B.}~\bibnamefont {Kryzhanovsky}}, \bibinfo {author} {\bibfnamefont {L.}~\bibnamefont {Litinskii}}, \ and\ \bibinfo {author} {\bibfnamefont {V.}~\bibnamefont {Egorov}},\ }\bibfield  {title} {\enquote {\bibinfo {title} {Analytical expressions for ising models on high dimensional lattices},}\ }\href {\doibase https://doi.org/10.3390/e23121665} {\bibfield  {journal} {\bibinfo  {journal} {Entropy}\ }\textbf {\bibinfo {volume} {23}},\ \bibinfo {pages} {1665} (\bibinfo {year} {2021})}\BibitemShut {NoStop}%
\bibitem [{\citenamefont {Dagan}\ \emph {et~al.}(2021)\citenamefont {Dagan}, \citenamefont {Daskalakis}, \citenamefont {Dikkala},\ and\ \citenamefont {Kandiros}}]{dagan2021learning}%
  \BibitemOpen
  \bibfield  {author} {\bibinfo {author} {\bibfnamefont {Y.}~\bibnamefont {Dagan}}, \bibinfo {author} {\bibfnamefont {C.}~\bibnamefont {Daskalakis}}, \bibinfo {author} {\bibfnamefont {N.}~\bibnamefont {Dikkala}}, \ and\ \bibinfo {author} {\bibfnamefont {A.~V.}\ \bibnamefont {Kandiros}},\ }\bibfield  {title} {\enquote {\bibinfo {title} {Learning ising models from one or multiple samples},}\ }in\ \href {\doibase https://doi.org/10.1145/3406325.3451074} {\emph {\bibinfo {booktitle} {Proceedings of the 53rd Annual ACM SIGACT Symposium on Theory of Computing}}}\ (\bibinfo {year} {2021})\ pp.\ \bibinfo {pages} {161--168}\BibitemShut {NoStop}%
\end{thebibliography}%

\end{document}